\documentclass[twocolumn]{aastex631}

\shorttitle{ISM Conditions in a Dusty, Low Mass, z=4.27 Main Sequence Galaxy}
\shortauthors{Mizener et al.}

\graphicspath{{./}{figures/}}

\begin{document}

\title{First Constraints on the ISM Conditions of a Low Mass, Highly Obscured z=4.27 Main Sequence Galaxy}

\author[0000-0003-2722-6600]{Andrew Mizener}
\affiliation{Department of Astronomy, University of Massachusetts Amherst MA 01003, USA: amizener@umass.edu}

\author[0000-0001-8592-2706]{Alexandra Pope}
\affiliation{Department of Astronomy, University of Massachusetts Amherst MA 01003, USA: amizener@umass.edu}

\author[0000-0002-6149-8178]{Jed McKinney}
\affiliation{Department of Astronomy, The University of Texas at Austin, 2515 Speedway Blvd Stop C1400, Austin, TX 78712, USA}

\author[0000-0001-9394-6732]{Patrick Kamieneski}
\affiliation{School of Earth and Space Exploration, Arizona State University, 
PO Box 871404,
Tempe, AZ 85287-1404, USA}

\author[0000-0001-7160-3632]{Katherine E. Whitaker}
\affiliation{Department of Astronomy, University of Massachusetts, Amherst, MA 01003, USA}
\affiliation{Cosmic Dawn Center (DAWN), Denmark}

\author[0000-0003-4569-2285]{Andrew Battisti}
\affiliation{Research School of Astronomy and Astrophysics, Australian National University, Cotter Road, Weston Creek, ACT 2611, Australia}

\author[0000-0001-7089-7325]{Eric Murphy}
\affiliation{National Radio Astronomy Observatory, 520 Edgemont Road, Charlottesville, VA 22903, USA}

\accepted{in ApJ May 7, 2024}

\begin{abstract}

We present the molecular gas content and ISM conditions of MACSJ0717\_Az9, a strong gravitationally lensed $z=4.273$, $M_{*} \simeq 2\times10^9M_{\odot}$ star-forming galaxy with an unusually high ($\sim 80\%$) obscured star formation fraction. We detect CO(4-3) in two independent lensed images, as well as [N II]205$\mu$m, with ALMA. We derive a molecular gas mass of log$_{10}[M_{H_{2}} (M_{\odot})] = 9.77$ making it moderately deficient in molecular gas compared to the lower redshift gas fraction scaling relation. Leveraging photodissociation region (PDR) models, we combine our CO(4-3) measurements with existing measurements of the [C II] 158$\mu$m line and total infrared luminosity to model the PDR conditions. We find PDR conditions similar to local star-forming galaxies, with a mean hydrogen density log$_{10}$[$n_H$ $cm^{-3}$] = $4.80\pm0.39$ and a mean radiation field strength log$_{10}$[G$_0$ Habing] = $2.83\pm0.26$. Based on Band 3 continuum data, we derive an upper limit on the intrinsic dust mass of log$_{10}[M_{\rm dust} (M_{\odot})] < 7.73$, consistent with existing estimates. We use the 3D tilted-ring model fitting code 3D-Barolo to determine the kinematic properties of the CO(4-3) emitting gas. We find that it is rotationally dominated, with a $V/\sigma=4.6 \pm 1.7$, consistent with the kinematics of the [C II]. With PDR conditions remarkably similar to normal dusty star-forming galaxies at z \textless 0.2 and a stable molecular disk, our observations of Az9 suggest that the dust-obscured phase for a low-mass galaxy at z$\sim$4 is relatively long. Thus, Az9 may be representative of a more widespread population that has been missed due to insufficiently deep existing millimeter surveys. 

\end{abstract}


\section{Introduction} \label{sec:intro}

In the late 1990s, the first sub-millimeter surveys with SCUBA \citep[e.g.][]{Smail1997,Barger1998} revealed a new population of massive, intensely IR-luminous dusty star-forming galaxies at high redshift. Though massive, dusty, infrared-luminous galaxies have been a popular subject of study in the intervening decades, our understanding of their highly-obscured analogues at lower masses, infrared luminosities, and star formation rates is limited. These systems are widespread: at $z > 2.5$, star-forming galaxies with stellar masses $M < 10^{10} M_\odot$ dominate the galaxy population by number count \cite[][]{Somerville2015,Solimano2021}. The subset of these galaxies with total stellar mass around $\sim 10^9 M_{\odot}$ at $z\sim4$ are the likely progenitors of galaxies with stellar mass similar to our Milky Way at $z\sim0$ \citep{Moster2013, vanDokkum2013}. However, due to observational challenges, they are not well studied at millimeter wavelengths. Probing the gas content, kinematics, and ISM conditions of these systems is vital to developing an understanding of how ``normal" galaxies assemble their mass and build up their dust and metal content in the early universe.

With a stellar mass of $2.14^{+1.04}_{-0.05} \times 10^{9} M_{\odot}$ and a total star formation rate of SFR$_{total} = 30.3$ $M_{\odot}$/yr, MACSJ0717\_Az9 (hereafter Az9) lies on the star-forming main sequence at its stellar mass and redshift \citep{Pope2017,Pope2023}. Az9 is very dust obscured - its obscured star formation rate (SFR) fraction is $0.83 \pm 0.12$, $\sim4\times$ the expected value at its stellar mass based on UV-selected galaxies \citep{Pope2017,Pope2023}. Taken together, the mass-metallicity relation \cite[MZR,][]{Tremonti2004, Sanders2021} and observations of the dust-obscured star formation rate at z$<$2.5 \citep{Whitaker2017} suggest that systems like Az9 should not be this heavily obscured. High-redshift objects have had less time to produce the metals that condense into dust \citep{Popping2017}, and lower-mass systems tend to be less obscured \citep{Whitaker2017} than their more massive counterparts. This begs the question: is Az9 a typical source for its mass and redshift? And how did it end up with such a high obscuration fraction? 

One way to tackle this issue is through the lens of PDR conditions. The dominant mechanism for dust production at high redshift is direct condensation in the ISM \citep{Michalowski2015,Popping2017}, the rate of which depends in part on the density and temperature of the medium \citep{Hirashita2000,Asano2013}. Stemming from the seminal work of \citet{Tielens1985}, a series of theoretical models of photodissociation regions (PDRs) have been produced allowing observers to link observed line ratios to the physical conditions (i.e. density, incident radiation field strength, and surface temperature) of PDRs \cite[e.g.][]{Wolfire1990,Kaufman1999,Kaufman2006}. Models like these have been used to describe the PDR conditions in a range of galaxies, from nearby normal dusty galaxies \citep[e.g.][]{Malhotra2001,Hughes2017} to IR-bright dusty massive galaxies at high redshift \cite[e.g.][]{Bothwell2017,Gullberg2015}. However, since observationally expensive tracers of the dense gas like [O I] or CO are required to constrain these models, the PDR conditions in low-mass systems at high redshift have not yet been extensively studied. 

The total molecular gas content of Az9 is also a part of the puzzle, as it is both the fuel for star formation and the environment in which dust grains condense. Observations of massive systems suggest that as we push to higher redshifts, galaxies tend to have larger molecular gas mass to stellar mass fractions (hereafter referred to as $\mu_{gas} =$ log$_{10}[{M_{H_2}} / M_{*}]$) compared to galaxies in the local universe \citep{Tacconi2010} at the same stellar mass. Additionally, systems with higher stellar masses are generally associated with lower $\mu_{gas}$\cite[][]{Tacconi2018}. Thus, we may naively expect low-mass high-redshift systems to possess very high $\mu_{gas}$ even relative to more massive systems at the same redshift. However, observations suggest that these systems possess $\mu_{gas}$ smaller than what scaling relations lead us to expect. Molecular gas mass to stellar mass fractions in the range $\mu \sim $ 0.2 to 0.5 are found in the few low-mass ($M_{*} \lessapprox 10^{10} M_{\odot})$ high-redshift $(z \gtrapprox 2.5)$ galaxies that have been observed \cite[e.g.][]{Saintonge2013,DZ2017,Solimano2021} compared to expected (via extrapolation) fractions around $\mu_{gas}$ $\sim$ 0.4 to 0.6 \citep{Tacconi2018} with expected scatter $\sim 0.2$ dex.

The mass assembly of galaxies is dominated by two processes: steady cold accretion \citep{Keres2005} and mergers \citep{Hopkins2010}. While steady cold accretion tends to produce systems with stable disks \citep{Forster2020,Rizzo2023}, mergers and counter-rotating gas inflows may disrupt galactic structure. Generally speaking, star-forming galaxies (SFGs) are less likely to host rotation-dominated disks as we push to lower stellar masses and earlier epochs \citep{Forster2020}. This is an unsurprising result; low-mass systems are easier to disrupt, and the harsher conditions of the early universe increases the efficacy of disruptive processes \citep[e.g.][]{Pillepich2019}. However, a number of dynamically cold disks have still been observed in the high-redshift regime \citep[e.g.][]{Rizzo2020, Rizzo2021, Jones2021b, Rizzo2022, Rizzo2023, Roman-Oliveira2023}.

Recent work with the James Webb Space Telescope (JWST) has identified a population of relatively low-mass dusty galaxies at $z < 6$, some of which were originally misidentified as $z > 10$ galaxies \citep[e.g.][]{Naidu2022,Barrufet2023,Nelson2023}. These systems, often referred to as ``HST-dark galaxies", likely represent a lower mass and SFR extension of the high redshift LIRG/ULIRG population \citep{Barrufet2023}. Such galaxies are likely missing from existing HST and Spitzer-selected samples, as well as single-dish submillimeter studies \citep{Barrufet2023}. As a highly obscured galaxy on the star-forming main sequence at moderately high redshift, Az9 presents a unique opportunity to study a potential member of this newly discovered population in great detail.

Intrinsically dim galaxies such as these HST-dark systems can be difficult to study without significant investments of telescope time. However, we can more easily access a subset of these systems (including Az9) by leveraging strong gravitational lensing. Depending on the distribution of matter producing such a field (the ``lens"), as well as the geometry between the source, lens, and observer, this phenomenon can increase the apparent angular area subtended by a source \citep{Wambsganss1998,Peacock1999}. Because gravitational lensing does not affect surface brightness \citep{Peacock1999}, this phenomenon amplifies lensed sources.

In this paper, we aim to characterize the ISM conditions and molecular gas content of Az9. Our paper is organized as follows. In Section 2 we describe our new ALMA CO(4-3) and [N II]205$\mu$m observations and introduce existing measurements of Az9 from the literature. Our analysis is found in Section 3, including our PDR modeling methodology, our kinematic modeling process and how we convert our CO measurements into a molecular gas mass. We present our results in Section 4, including total molecular gas mass, dynamical mass, metallicity, and PDR conditions. We finish with our discussion in Section 5. Throughout this paper we assume a flat $\Lambda CDM$ cosmology with $H_0 = 69.6$ km s$^{-1}$ Mpc$^{-1}$ and a Chabrier IMF \citep{Chabrier2003}.

\section{Observations \& Data Reduction} \label{sec:obs}

\begin{figure}[htb!]
\includegraphics[width=0.45\textwidth]{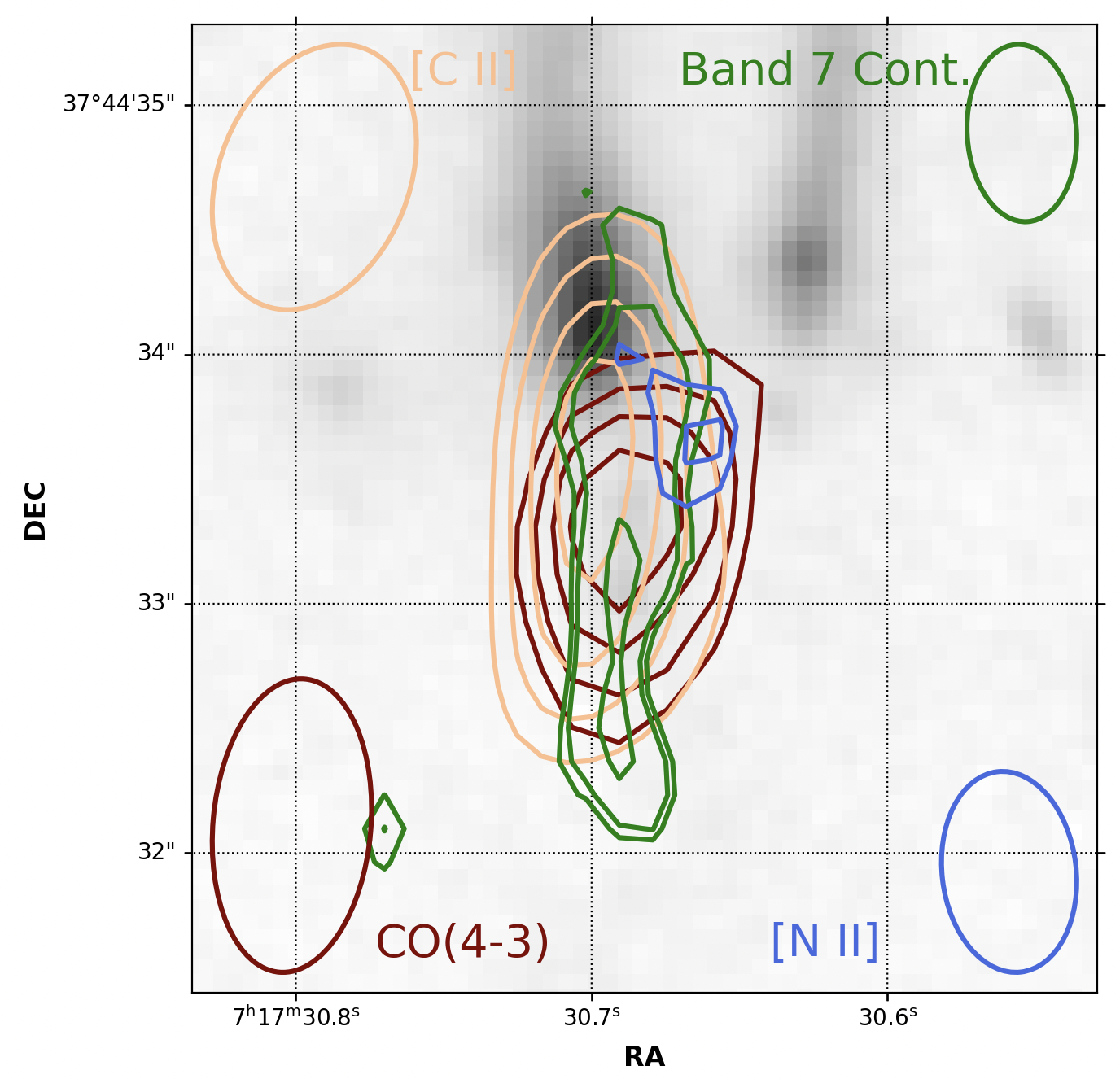}
\caption{A HST F160W (rest-frame UV) cutout of the Az9 (5.2) image overlaid with CO (red), [C II] (yellow), [N II]205$\mu$m (blue), and Band 7 continuum (green) contours. The beams corresponding to each ALMA observation are also provided in the corners. The CO and [C II] contours are shown at 4, 5, 6, and 7$\sigma$. Due to lower signal-to-noise, the [N II] and Band 7 continuum contours are shown at 2.5, 3.0, and 4.5 $\sigma$.}
\label{fig:map}
\end{figure}

\subsection{Target}
The cluster MACSJ0717.5+37.45 was imaged with the Hubble Space Telescope (HST) as part of the Hubble Frontier Fields (HFF) program. Lensing models of this cluster consistently identify a multiply lensed system at z$\sim$4 with three lensed components in the image plane \citep{Limousin2016,Diego2015,Johnson2014,Zitrin2015}. Following the naming convention from \cite{Limousin2016} we call these components 5.1, 5.2, and 5.3. It is
possible for a lensing configuration like this to have one image containing the full galaxy and a pair
of images (with reverse parity) merging together at a line of symmetry. The 5.2 arc studied in this paper
is part of a pair of merging images, its counterpart being the 5.1 image. In some cases, each member of a merging pair may not be representative of the lensed galaxy as a whole. However, in our case, we are confident we are recovering flux from a complete image of Az9 (see Section 3.4). By using existing HST, Spitzer, Large Millimeter Telescope (LMT), and Atacama Large Millimeter Array (ALMA) data \citet{Pope2023} obtained a stellar mass, star formation rate, and total IR luminosity for 5.2 (hereafter Az9). The [C II] line emission from Az9 has also been measured with ALMA \cite[][]{Pope2023}.

\subsection{Band 3 Observations}
{The CO(4-3) line ($\nu_{rest} = 461.04$ GHz) in MACSJ0717 Az9 was observed over the course of 6 nights in December 2021 with the Atacama Large Millimeter Array (ALMA) in Band 3 as a part of proposal 2021.1.00272.S (PI: Pope). The array spent time in the C-6, C-5, and C-4 configurations over the course of these observing sessions, resulting in a nominal configuration of C43-5. The primary beam (HPBW $\simeq 66^{\prime\prime}$) was centered on the 5.2 image of Az9, but also covered 5.1 which is $\sim 17^{\prime\prime}$ away. Total on source observing time was 8.67 hours. Precipitable water vapor (PWV) varied from 3.3mm to 8.3mm over the course of these observations.

The data are calibrated with the standard ALMA pipeline (Version 2021.2.0.128), which utilizes the Common Astronomy Software Applications (CASA) package \cite[][Version 6.2.1.7]{CASA}. We search for both continuum and line emission in these data using the CASA routine \textit{tclean}. We determined that the JvM correction is not required for this source, as in these data the clean beam is a very good approximation to the shape and volume of the dirty beam on scales containing real emission.

\subsubsection{Continuum}
Because the presence of strong continuum emission can potentially effect measured line characteristics, we first use \textit{tclean} in continuum mode (excluding channels expected to contain line emission) to produce a continuum image. After interactive cleaning, we find that the continuum emission is undetected allowing us to place an upper limit on continuum emission from the image RMS of $\sigma = 0.004$ mJy/beam.

Although Az9 is expected to be dust-rich, a non-detection of the continuum emission is still consistent with existing model SEDs (see Section 3.3). With a beam size of $1.165^{\prime\prime} \times 0.629^{\prime\prime}$ with natural weighting and using the same aperture used for the CO, this corresponds to a $1\sigma$ upper limit on the continuum flux of 0.015 mJy at the frequency we measure the continuum, 94.3 GHz (i.e. $600 \mu$m).

\subsubsection{Line}
To image the line emission, use a standard Hogbom deconvolver with Briggs weighting and use cell sizes sufficient to place 6 pixels across the restoring beam major axis for each clean attempt. We image the whole FOV, capturing both the 5.1 and 5.2 lensed images. Re-imaging with the 5.1 image at the phase center did not substantially improve its detection, so we consistently use the map with 5.2 at the phase center such that all reported values are derived from the same CLEAN map. We produce cubes using an array of different Briggs weights ranging from 0.0 to 2.0 with an 0.5 step in order to explore how various weighting schemes affect our results. Inspecting each cube leads us to select the Briggs 1.0 cube as our final science image, as this weight produces a relatively fine resolving beam that does not miss significant observed line flux when compared to natural weighting.

This final image cube has a per-channel rms of $0.052$ mJy / beam with a channel velocity width of $53.6$ km/s, the native channel width of the observations. The maximum recoverable scale for ALMA at Band 3 in the C-6 configuration is $4.11^{\prime\prime}$. As the largest scale across which we measured emission is $\sim1.5^{\prime\prime}$, our observations do not resolve out any significant emission. At the location of the 5.1 image, the sensitivity is $\sim 86\%$ of the sensitivity at the 5.2 image due to primary beam falloff.

\begin{rotatetable*}
\movetableright=0.5mm
\begin{deluxetable*}{lcccccc||ccc}
\tablewidth{0pt}
\tablehead{
\colhead{Component ID} &
\colhead{RA} &
\colhead{DEC} &
\colhead{z} &
\colhead{$S_{cont}$} &
\colhead{$\Delta v$ (FWHM)} &
\colhead{$S_{line} \Delta v$} &
\colhead{$L^\prime_{CO(4-3)}$} &
\colhead{$L_{CO(4-3)}$} &
\colhead{Magnification}\\
\colhead{ID} &
\colhead{J2000} &
\colhead{J2000} &
\colhead{} &
\colhead{mJy} &
\colhead{(km/s)} &
\colhead{$10^{-1}$ (Jy km/s)} &
\colhead{$\mu$}}
\tablecaption{MACSJ0717 Az9: Observed Quantities}
\tablewidth{0pt}
\tablehead{
\colhead{Component ID} &
\colhead{RA} &
\colhead{DEC} &
\colhead{z} &
\colhead{$S_{\text{cont}}$} &
\colhead{$\Delta v$ (FWHM)} &
\colhead{$S_{\text{line}} \Delta v$} &
\colhead{Magnification} &
\colhead{$L^\prime_{\text{CO(4-3)}}$} &
\colhead{$L_{\text{CO(4-3)}}$} \\
\colhead{ID} &
\colhead{J2000} &
\colhead{J2000} &
\colhead{} &
\colhead{mJy} &
\colhead{(km/s)} &
\colhead{$10^{-1}$ (Jy km/s)} &
\colhead{$\mu$} &
\colhead{$10^{9}$ (K km/s)} &
\colhead{$10^{6}$ ($L_{\odot}$)}}
\startdata
\multicolumn{1}{c}{\textit{CO(4-3)}} & \multicolumn{6}{c||}{} & \multicolumn{3}{c}{} \\
\multicolumn{1}{c}{\textit{($\lambda_{\rm rest} = 650\mu$m)}} & \multicolumn{6}{c||}{} & \multicolumn{3}{c}{} \\
\textbf{5.2} & 07:17:30.71 & +37:44:33.33 & $4.2734 \pm 0.0002$ & $<0.046$ & $250 \pm 22$ & $1.80 \pm 0.27$ & $7.5 \pm 1.0$ & $1.21 \pm 0.18$ & $3.54 \pm 0.53$ \\
\textbf{5.1} & 07:17:31.18 & +37:44:48.70 & $4.2735 \pm 0.0003$ & $<0.044$ & $305 \pm 50$ & $1.63 \pm 0.46$ & $6.8 \pm 1.1$ & $1.09 \pm 0.31$ & $3.20 \pm 0.90$ \\
\multicolumn{1}{c}{\textit{[N II]205$\mu$m}} & \multicolumn{6}{c||}{} & \multicolumn{3}{c}{} \\
\multicolumn{1}{c}{\textit{($\lambda_{\rm rest} = 205\mu$m)}} & \multicolumn{6}{c||}{} & \multicolumn{3}{c}{} \\
\textbf{5.2}& 07:17:30.74 & +37:44:33.60 & $4.2702 \pm 0.0002$ & $1.03 \pm 0.16$ & $150 \pm 30$ & $2.36 \pm 1.02$ & $7.5 \pm 1.0$ & &
\enddata
\tablecomments{Although both images have a similar total line flux, the detection of image 5.1 is less significant because it lies further from the primary beam center, where the RMS level is higher. $S_{\text{cont}}$ upper limits are at the 3$\sigma$ level. Values to the left of the double line have not been corrected for lensing magnification. Luminosity value to the right of the double line have been corrected for lensing magnification. A primary beam correction has been applied to the values shown here.}
\end{deluxetable*}
\end{rotatetable*}

\subsection{Band 7 Observations}
Az9 was also observed in Band 7 during a single night (2022 September 14) with ALMA as part of proposal 2021.1.00272.S (PI: Pope) with the intention of measuring the [N II]205$\mu$m line ($\nu_{rest} = 1461.13$ GHz). The total time on source was 26 minutes, only 25$\%$ of what was requested, with 0.24mm PWV. The array was in the C43-3 configuration during these observations, resulting in a maximum recoverable scale of 5.5 $^{\prime\prime}$.

As with the CO data, these  data are calibrated with the standard ALMA pipeline (Version 2021.2.0.128). Because these observations do not achieve the requested depth, we use natural weighting to maximize our sensitivity for both continuum and line imaging.

\subsubsection{Continuum}
After cleaning in multifrequency synthesis mode, continuum flux is extracted within an elliptical aperture and the uncertainty is determined following the procedure used in \cite{Pope2023}. The final continuum image has rms of $0.06$ mJy / beam and beam size 0.71$^{\prime\prime}$ $\times$ 0.44$^{\prime\prime}$.

\subsubsection{Line}
We find relatively strong continuum emission in this band, so we first perform continuum subtraction using \textit{uvcontsub}. We fit a 1st-order polynomial over the range 277.763 GHz - 277.279 GHz and 277.201 GHz - 275.919 GHz, corresponding to all unmasked channels not containing line emission. After cleaning in cube mode, the final line image has a per-channel rms of $0.66$ mJy / beam and beam size 0.81$^{\prime\prime}$ $\times$ 0.54$^{\prime\prime}$.

\begin{figure*}[!htb]
\gridline{\fig{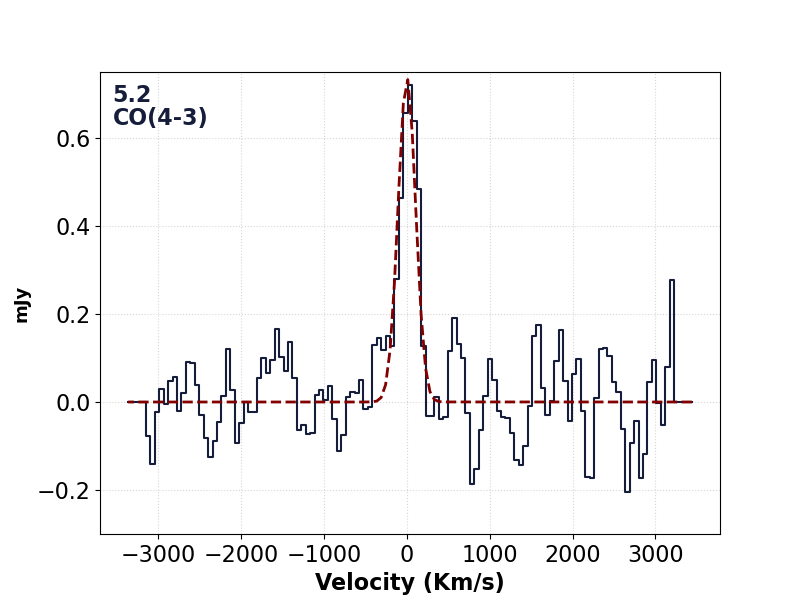}{0.5\textwidth}{(a)}
          \fig{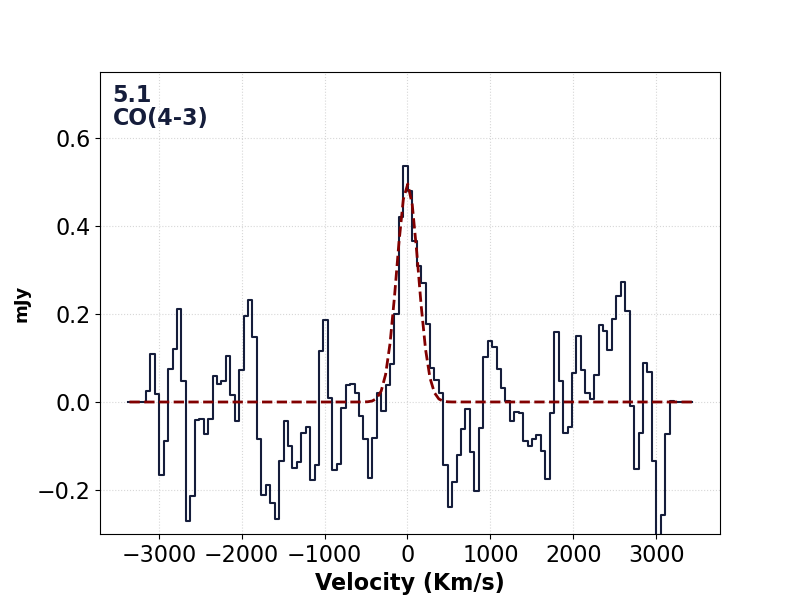}{0.5\textwidth}{(b)}}
\gridline{\fig{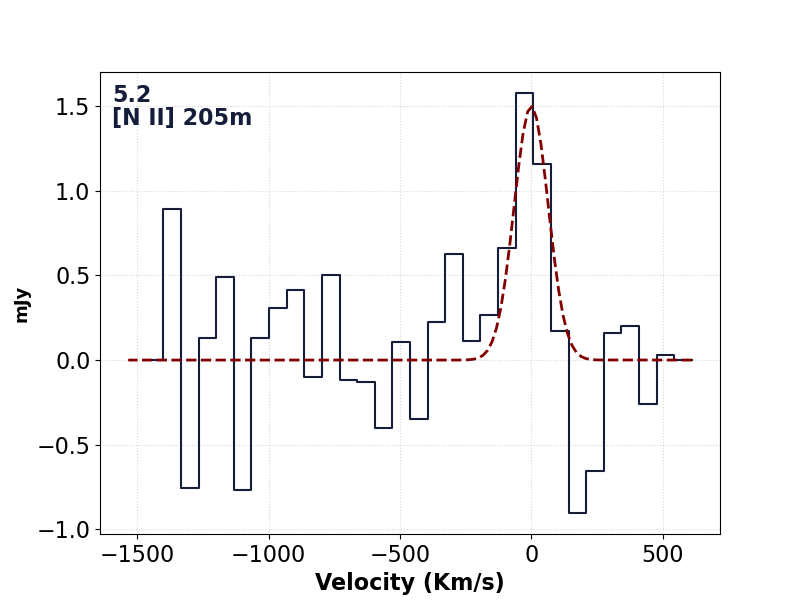}{0.5\textwidth}{(c)}}  
\caption{Observed (i.e. not lensing corrected) spectra of Az9; CO(4-3) 5.2 (a), 5.1 (b) and [N II] 5.2 (c). Each spectrum is overlaid with the best-fit Gaussian. For visualization purposes, the [N II] spectrum has been binned by a factor of 4. A primary beam correction has been applied to each spectrum. Given their relative magnifications, the CO lines are in agreement for 5.2 and 5.1. In these plots, we set the zero velocity position to the location of each line center.}
\label{fig:spec}
\end{figure*}

\section{Analysis} \label{sec:analysis}

\subsection{CO(4-3) and [N II]205$\mu$m Measurements}

We perform our spectral extraction in the image (i.e. unreconstructed) plane. We perform a Gaussian fit to the extracted spectrum of each lensed image within a custom aperture (used to optimally include low-surface-brightness, $\sim 2\sigma$ emission near the edge of the source) using the \cite{astropy1} software package. The CO emission is well modeled by a Gaussian in both lensed images, with a reduced chi-squared of 0.638 for the 5.2 component and 0.522 for the 5.1 component (Figure 2). The [N II]205$\mu$m emission in 5.2 is also modeled with a Gaussian with a reduced chi-squared of 1.41. The CO(4-3) detection is significant in both lensed images; $\sim8.1\sigma$ and $\sim4.7\sigma$ in the 5.2 and 5.1 images respectively. Binning the spectrum by a factor of 2 did not have a significant impact on the derived line intensity. The [N II]205$\mu$m detection is less significant at $\sim2.4\sigma$; only the 5.2 image is within the FOV. The redshifts of each CO component agree with one another within their mutual uncertainties. The [N II]205$\mu$m is offset from the peak in the CO (5.2 image) by $\sim 184 \pm 25$ km/s blueward. However, compared to the CO velocity at the spatial location of the [N II]205$\mu$m} emission, the velocity offset between the two is only $\sim60 \pm 25$ km/s. Because the low SNR of the [N II]205$\mu$m observations, we refrain from making any physical interpretation of this offset until a higher quality [N II]205$\mu$m observation can be obtained.

We provide the observed CO(4-3) and [N II]205$\mu$m properties of Az9 in Table 1. In addition to our measurements of line flux, we also provide the CO line luminosity in units of $L_\odot$ and the areal integrated source brightness temperature in units K km/s pc$^2$ (often referred to as L$'_{CO(4-3)}$). For the remainder of our analysis, we use our measurement of the 5.2 image as it is a more reliable detection than 5.1. We also apply a lensing magnification correction when working with intrinsic quantities. For the remainder of this work, we will refer to lensing-corrected quantities as \textit{intrinsic} and uncorrected quantities as \textit{observed}.

\subsection{Total Dust Mass}
We place an upper limit on the dust mass of Az9 based on our Band 3 continuum ($600 \mu$m rest frame) observations. Although we detect continuum emission in our Band 7 observations \citep{Pope2023}, this does not correspond to emission from the Rayleigh-Jeans tail of the rest-frame dust SED and so cannot be used to constrain the total dust content. To produce a dust mass upper limit, we use the prescription given in Section 3.4 of \cite{Solimano2021}, setting our dust temperature to $T_{dust} = 25$K and using a standard dust emissivity index of $\beta$=1.8. This $T_{dust} = 25$K is the expected temperature of the cold dust component that dominates the dust mass \citep{Scoville2016,Kaasinen2019}, this is not the same as the luminosity-weighted dust temperature, which cannot be used to estimate a dust mass. 

Given our continuum observations are centered at a rest wavelength of $600\mu$m, we use a dust mass absorption coefficient of 0.9 cm$^2$ g$^{-1}$ \citep{Draine2001}. With these parameters, we derive an intrinsic $3\sigma$ dust mass upper limit of log$_{10}[M_{dust} (M_{\odot})] < 7.73$ after correcting for lensing magnification. This dust mass is consistent with expectations from the MAGPHYS SED: log$_{10}[M_{dust,SED}] = 7.55^{+0.20}_{-0.13}$ \citep{Pope2023}. The dust mass estimate from MAGPHYS is less reliable than our new upper limit, as it is constrained by data from near the peak of the dust SED rather than the Rayleigh-Jeans tail (i.e. it traced the warm rather than cold dust component).

We note that there is some debate in the literature as to whether the mass-weighted dust temperature is in fact equivalent to the luminosity-weighted dust temperature in some systems \citep[e.g.][]{Harrington2021}. If we assume a luminosity-weighted dust temperature based on the SED fits of \citet{Pope2023} ($T_{dust} = 40$K), we derive a dust mass upper limit of log$_{10}[M_{dust} (M_{\odot})] < 7.52$, also consistent with the mass estimate from those SED fits.

\subsection{CO Conversion Factors}

It is standard to estimate molecular gas masses from higher-J transitions by assuming a ratio of CO(J-(J-1)) to CO(1-0) \citep{Carilli2013} since the CO(1-0) transition in high-redshift sources is often intrinsically faint and may fall into unsuitable atmospheric windows. One of the main sources of systematic uncertainty in our analysis is this conversion from CO(4-3) to CO(1-0), $r_{41}$. Recently, \cite{Castillo2023} found a range of $r_{41} = 0.63 \pm 0.44$ in a sample of 30 massive, high redshift, star-forming galaxies. For the molecular gas mass, our PDR modeling and our kinematic analysis, we use the \citet[][]{Castillo2023} distribution.

CO line measurements are used to infer molecular gas masses via a quantity called $\alpha_{CO}$, which converts a CO(1-0) areal integrated source brightness temperature into a total molecular gas mass. A range of $\alpha_{CO}$ values have been reported in the literature, from $\alpha_{CO}=$ 0.8 $M_{\odot}$ / K km s$^{-1}$ pc$^2$ in submillimeter galaxies (SMGs) and quasars up to $\alpha_{CO}=$ 4.8 $M_{\odot}$ / K km s$^{-1}$ pc$^2$ in color-selected galaxies (CSGs) \citep{Carilli2013}. $\alpha_{CO}$ can be significantly higher in some environments - values as high as $\sim 70$ have been inferred in the SMC \cite[][]{Leroy2011}. In general, the conversion factor is expected to increase as metallicity decreases because not only are fewer CO molecules able to form in low-metallicity environments, but also because CO is photodissociated to larger depths within low-metallicity clouds \citep{Bolatto2013}. The typical approach is to rely on known scaling relations in order to infer the $\alpha_{CO}$ in a given system or to adopt a reasonable $\alpha_{CO}$ from the literature. The assumption of a fixed gas-to-dust ratio is also sometimes used. As CO is generally optically thick, its luminosity depends most strongly on the physical size and velocity dispersion of the emitting cloud \citep{Bolatto2013}. This produces a higher $\alpha_{CO}$ in low-metallicity environments, which possess smaller CO emitting regions relative to the $H_2$ cloud size. Dust is also expected to play an important role as dustier environments are better at shielding CO from photodissociation.

We use a flat distribution of $\alpha_{CO}$ with an upper limit motivated by the kinematics of Az9 (as discussed in Section 4.4.1) and a lower limit corresponding to the QSO value from \citet{Carilli2013}. Additionally, we provide constraints on  metallicity and discuss the implications that these constraints gave on the $\alpha_{CO}$ of Az9 in Section 4.5.

\subsection{Source Plane Reconstruction}

\begin{figure}[htb!]
\plotone{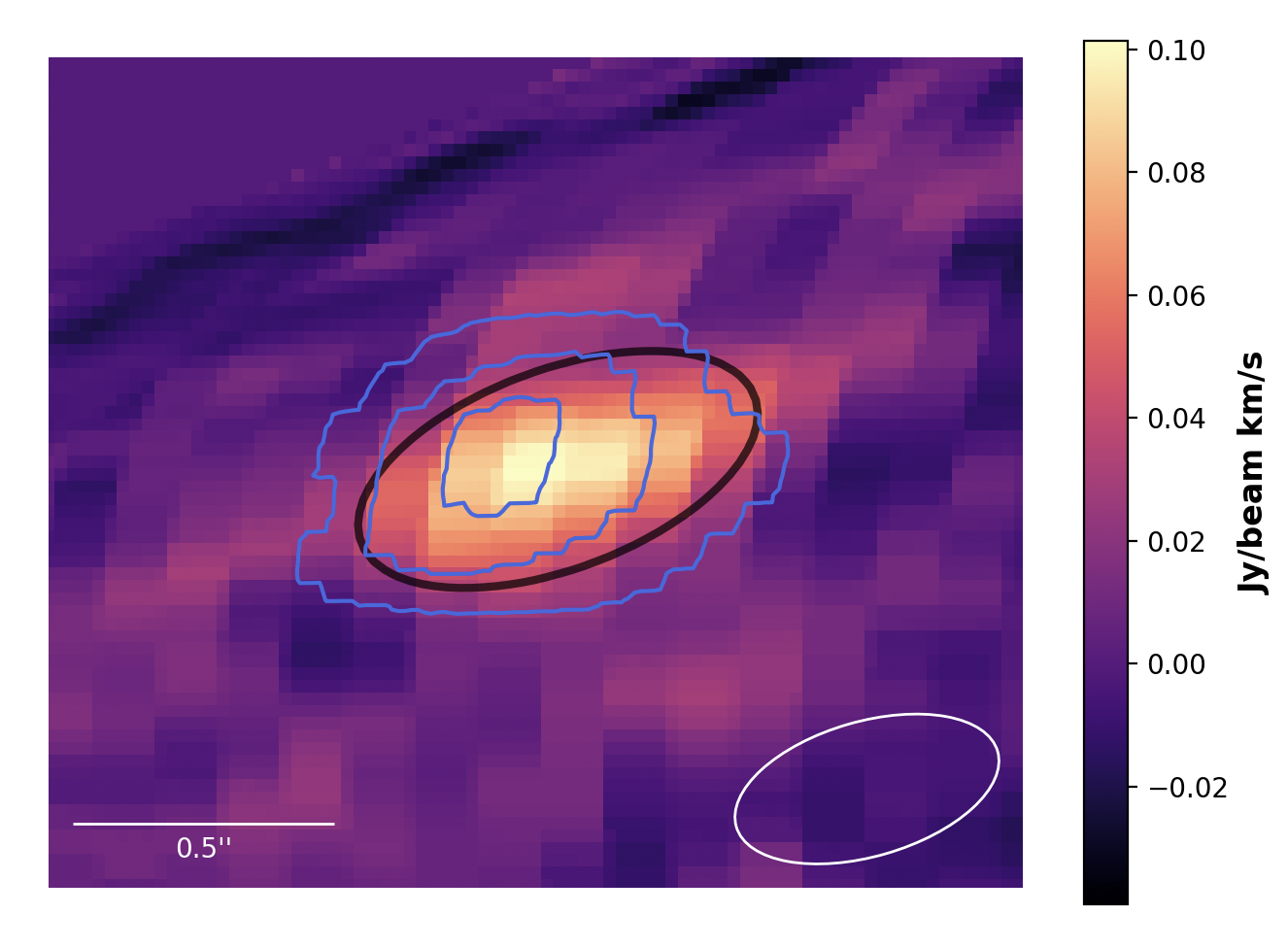}
\caption{Our moment 0 source-plane reconstruction of the CO(4-3) emission from Az9 (image 5.2). Contours of the source-plane [C II] emission are placed at the 3, 6, and 9$\sigma$ levels in blue. We place a single black contour at As9's effective radius obtained via Sersic fitting. The white ellipse shows the representative source-plane reconstructed synthesized beam.}
\label{fig:recon}
\end{figure}

From the public HFF lens models,\footnote{https://archive.stsci.edu/prepds/frontier/lensmodels/} we use the non-cored model produced by \citet{Limousin2016}. We find that the relative CO fluxes of the 5.1 and 5.2 images are consistent with each other given the \cite{Limousin2016} model prediction.  Uncertainties in the lensing correction are propagated through our gas mass and PDR analysis.

We perform a source-plane reconstruction at our observed spectroscopic CO(4-3) redshift of $z=4.27$ with Lenstool \cite[][]{Kneib1996,Jullo2007,Jullo2009}. We apply this reconstruction on a channel-by-channel basis. We also produce an approximate channel-averaged reconstructed beam by placing the image-plane synthesized beam in the spectral and spatial center of the image. As the magnification across any individual image of Az9 is relatively constant, this produces a reasonable approximation to the true source-plane beam. Because the 5.1 image is further from the primary beam center in our data, our source-plane reconstruction of it is of lower quality than 5.2. We thus elect to perform all source-plane analysis (i.e. kinematic analysis) on the 5.2 image. The 5.2 image is 0.75$^{\prime\prime}$ away from the caustic curve in the minor axis direction in the source plane. Given this distance and the Sersic fit described below, at most $\sim10\%$ of the flux from the 5.2 could be lost beyond the caustic, which is well within the uncertainty in line flux.

To help us further understand the nature of Az9 in the source plane, we perform a 2D Sersic profile fit \cite[][]{astropy1}. We show our reconstructed source-plane image in Figure 3. We over-plot a contour corresponding to the flux level at the half-light radius obtained via 2D Sersic fitting. The results of the Sersic fit are described in Section 4.1.

\subsection{PDR Modeling}
In order to model the physical conditions within PDRs, we use the PDR models of \citet{Kaufman1999,Kaufman2006} and \citet{Pound2008,Pound2023}. We note that a single PDR, as modeled here, is not a realistic description of the ISM of an entire galaxy. In practice, the ISM of any galaxy is composed of a mix of gas in multiple phases. This exercise provides a simplified picture that is only broadly representative of the PDR conditions in Az9. These models are constructed as infinite plane-parallel slabs of hydrogen illuminated on one side by an incident FUV radiation field. In these models, the gas is primarily heated via photoelectrons ejected from dust grains and PAH molecules. Gas cooling arising from line emission is handled by solving for chemical and energetic equilibrium in the slab under the escape probability formalism. Even though PDR models only formally apply to an individual star-forming region, the results when applied to a whole galaxy provide a luminosity-weighted average of the PDR conditions across the galaxy as a whole, and so should still be reliable if we wish to make broad comparisons between different galaxy populations.

We apply a few correction factors to our observed quantities. Because low-J CO emission is almost certainly optically thick, we multiply the observed CO luminosities by a factor of two to account for emission from shielded cloud regions following the procedure of \citet[][]{Hughes2017,Tadaki2019}, among others. [C II] and FIR emission are generally optically thin, so we do not apply any optical thickness correction to these luminosities. We note that specific approaches vary throughout the literature; for example, an equivalent correction (not applied in this work) is sometimes performed by keeping the luminosity of optically thick tracers the same and instead dividing the luminosity of optically thin tracers by a factor of 2 \citep{Hashimoto2023}.

Furthermore, we only want to use the [C II] emission arising from PDRs in this model, excluding the component arising from ionized regions. We use a relation derived by \citet{Croxall2017} (which requires only a measurement of the [C II] and [N II]205$\mu$m fluxes) to estimate the [C II] PDR fraction in Az9, which we use as a correction to the total [C II] luminosity in our PDR model. Via their equation in \citet{Croxall2017}, we find that the [C II] PDR fraction is $f_{neutral} = 0.83$. This is comparable to [C II] PDR fractions found in other galaxies across redshifts, which typically range from 0.5-0.85 \citep{Hughes2017}. Although there are some differences in approach from study to study, corrections like these are standard practice \citep[e.g.][]{Hughes2017,DiazSantos2017,Tadaki2019,Leung2019,Rybak2019,Ono2022,Hashimoto2023} and should be performed if we wish to fairly compare Az9 to other sources studied in the literature. For clarity, we note that our [N II]205$\mu$m measurement is not used in the PDR modeling except for this [C II] PDR fraction calculation.

We apply a Monte Carlo approach to our PDR modeling, drawing from a normal distribution centered on the measured value of each input quantity with $\sigma$ equal to the measured uncertainty in that quantity. We use two luminosity ratios as inputs to the model: [C II] / CO(1-0) and [C II] / FIR. In the context of this PDR model, the [N II]205$\mu$m is used only to estimate the [C II] PDR fraction as described in the previous paragraph. In order to understand the systematic effects caused by our unknown $r_{41}$, we use the \cite{Castillo2023} distribution $r_{41}=0.63 \pm 0.44$ as our CO(4-3) to CO(1-0) conversion factor. One run of our PDR modeling routine is performed with the $r_{41}$ fixed at 0.63; in another, we draw from a normal distribution like we do for the observed quantities. We perform 1000 runs of the model using each of these two approaches in order to fully sample the distribution of our uncertainties. The results of this analysis are described in Section 4.3.

\subsection{Kinematic Analysis}
We model the molecular gas kinematics of Az9 by applying 3D-Barolo, a 3-D tilted ring model fitting code \citep{Teodoro2015}, to our source-plane reconstruction of Az9 following the approach of \citet{Pope2023}. Using a forward-modeling approach, 3D-Barolo is able to robustly recover disc galaxy kinematic properties even at very low spatial resolution \citep{Teodoro2015,Teodoro2016,Rizzo2022}.

We begin by using 3D-Barolo's SEARCH module to isolate the regions in each channel containing real emission. We use a primary SNR cut of $4.5\sigma$, a secondary SNR cut of $3.5\sigma$, and we require regions to contain at least 30 pixels and contain significant flux in at least 2 consecutive channels to be included in the final mask. Once real emission has been isolated by SEARCH, we use 3D-Barolo's primary fitting module 3DFIT. This module simulates spectral cube observations by building a model made up from multiple concentric rings, which it combines to produce a model of a rotating gaseous disk which is compared to observations after convolution with the resolving beam.

To set the number and radius of rings used in our fit, we follow the approach of \cite{Jones2021b} and use a maximum model radius of $0.8\times$(the major axis FWHM of a 2D Gaussian fit to the de-convolved CO emission) and a minimum ring radius of the beam minor axis divided by 2.5. We use a minimum model radius of 0, and a number of rings determined by dividing the maximum model radius by the ring width. We use as initial guesses for the inclination and position angle the results from modeling of the [C II] emission \cite[][]{Pope2023}.

For $V_{max}$ and $V_{\sigma}$, we use initial guesses of 150 km/s and 30 km/s based on initial inspection of the moment 1 and moment 2 maps. We perform a preliminary run of 3D-Barolo with the ring centers, position angle, and inclination as free parameters to estimate their values. We then perform a second run with these parameters fixed at their means from the previous run, allowing only the rotational velocity and velocity dispersion to vary, in order to determine our final fitted values. We find an inclination of 47\textdegree.6 and a position angle of 188\textdegree.9 compared to an inclination of 46\textdegree.6 and a position angle of 189\textdegree.2 in the [C II] \cite[][]{Pope2023}. The results of our kinematic analysis are described in Section 4.4.  

\subsection{Molecular Gas Mass Uncertainties}
Accurately estimating a molecular gas mass from mid or high-J CO observations is a difficult business and must be done with care. The significant systematic uncertainties in $r_{41}$ and $\alpha_{CO}$ mean that the same CO observation could be interpreted as any gas mass in a relatively wide range, based on which fiducial values one selects for these parameters. As such, we wish to conservatively quantify the total uncertainty on our gas mass estimate including these systematics over the entire range of physically reasonable guesses.

We use a 5000-run Monte Carlo approach to explore how possible combinations of reasonable estimates for $r_{41}$ and $\alpha_{CO}$ impact the inferred molecular gas mass of Az9. We first draw the value of $L'_{CO(4-3)}$ from a normal distribution based on our observed $L'_{CO(4-3)}$, observed error, and lensing correction. Then, for each sample, we draw a value for $r_{41}$ and $\alpha_{CO}$ from distributions described in the next paragraph in order to estimate $M_{\rm H_2}$.

We note that there may be additional systematics that are not propagated through this exercise. For example, it is possible that at high redshift, more extreme excitation conditions cause a lower fraction of the total molecular gas mass to be traced by the CO(1-0) line than in the local universe, making our estimate too low. However, without observations of additional CO lines, we cannot definitively address this issue.

In this exercise, we use the $r_{41} = 0.63 \pm 0.44$ found by \citet[][]{Castillo2023}, excluding any negative values. We draw from a uniform $\alpha_{CO}$ distribution $\alpha_{CO} = 0.8 - 6.1 $, with the lower limit corresponding to the 0.8 value for SMGs and QSOs \citep{Carilli2013} and the upper limit motivated by our kinematic work in Section 4.4.1. As quantities such as the molecular gas fraction require a measure of stellar mass to be evaluated, we also draw stellar masses from a distribution based on the results and uncertainties of the MAGPHYS fit performed by \cite[][]{Pope2023}. The results of this analysis are described in Section 4.2.

\section{Results} \label{sec:results}

\subsection{Source-Plane Properties}
 After fitting a 2D Sersic profile to our source plane reconstruction of the CO(4-3) emission, we find that it has a best-fit half-light radius of 0.4$^{\prime\prime}$, an ellipticity of 0.52, and a Sersic index of 0.79. The CO is  more extended than the [C II], which had a half-light radius of 0.26$^{\prime\prime}$.

\subsection{Total Molecular Gas Content}
We show the intrinsic molecular gas content of Az9 compared to the \cite{Tacconi2018} relations and a sample of galaxies from the literature in Figure 4. We display this in terms of molecular gas mass to stellar mass fraction (defined as $\mu_{gas} = \log_{10}[M_{H_2}/M_{*}]$) in the upper panel, and in terms of the molecular gas depletion time (defined as $t_{dep} = [M_{H_2}/SFR]$) on the lower panel.

\begin{figure}[!htb]
\epsscale{2.3}
\plottwo{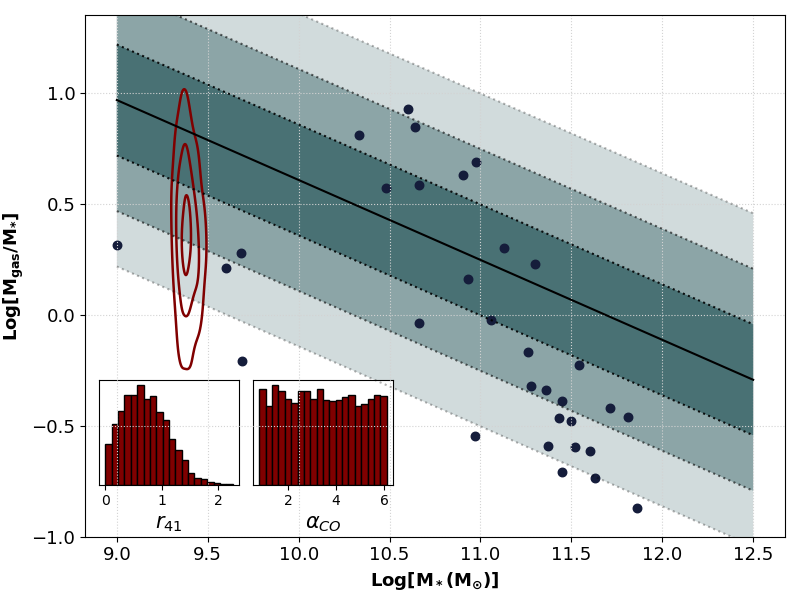}{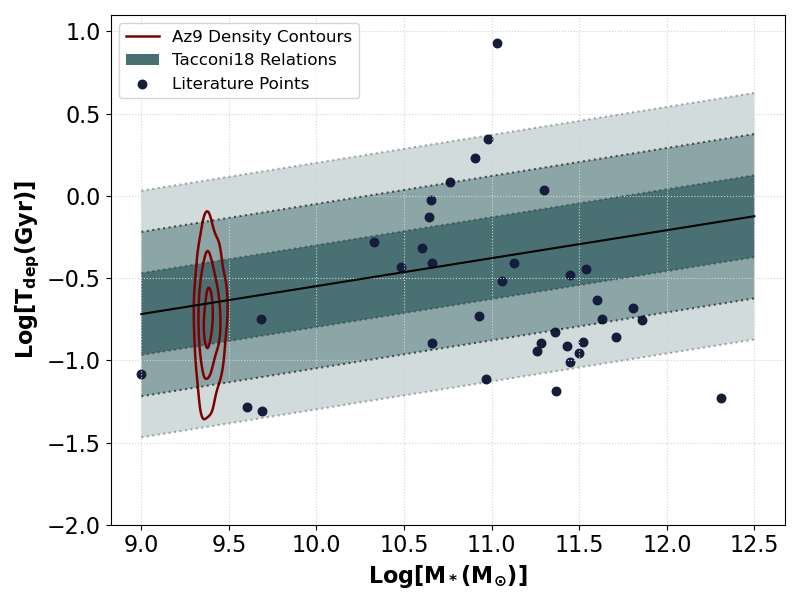}
\caption{Molecular gas mass to stellar mass fraction ($\mu_{M_{H_{2}}} = \log_{10}[M_{H_{2}}/M_{*}]$) (top) and depletion time ($t_{dep} = [M_{H_{2}}/SFR]$) (bottom) for Az9 compared to literature data and the \cite{Tacconi2018} relations evaluated at the redshift and $\delta_{MS}$ (distance from the star-forming main sequence) of Az9. Shaded regions are drawn at 1$\sigma$, 2$\sigma$, and 3$\sigma$ distance from the \cite{Tacconi2018} relations. The $r_{41}$ and $\alpha_{CO}$ distributions used for Az9 are shown in inset axes. Literature points in these plots were selected from systems with CO-based molecular gas measurements with $z \geq 2.5$, and were adjusted to represent a uniform $\alpha_{CO} = 3.6$ \cite[][]{DZ2017,Rivera2018,Magdis2012,Jones2021b,Birkin2021,Reichers2010a,Johansson2012,Saintonge2013}.
\label{fig:gasprop}}
\end{figure}

Contours are placed around the Monte Carlo results at the 0.3, 0.6, and 0.9 levels via a kernel density estimate. We derive a total molecular gas mass of log$_{10}[M_{H_{2}}(M_{\odot})] = 9.82 \pm 0.09$ using the fiducial $r_{41}$ and median $\alpha_{CO}$. This result implies a gas fraction of $\mu_{M_{H_2}} = 0.49 \pm 0.12$ and a depletion time of $0.22 \pm 0.05$ Gyr. The error we report on these quantities is derived from measurement uncertainties only. If the systematic uncertainties from $\alpha_{CO}$ and $r_{41}$ are included (as shown in Figure 4), we find that the median inferred gas mass is log$_{10}[M_{H_{2}}(M_{\odot})] = 9.77 \pm 0.43$. The small offset between the two mass estimates occurs because the fiducial $r_{41}$ is not the median of the distribution once non-physical negative $r_{41}$ draws are removed. In terms of its overall molecular gas content, Az9 appears to have a molecular gas mass to stellar mass fraction $\sim2.0\sigma$ below, and a depletion time consistent with, the main-sequence expectation at its stellar mass and redshift. However, given the substantial systematic uncertainties involved, we do not consider the offset in molecular gas content significant.

\begin{deluxetable}{ccc}


\tablecaption{Intrinsic ISM Properties of Az9}

\tablenum{2}

\tablehead{\colhead{Quantity} & \colhead{Unit} & \colhead{Value (Az9)} } 
\startdata
log$_{10}[M_{\rm dust}]$ & $M_{\odot}$ & $< 7.73$ \\
log$_{10}[M_{H_{2}}]$ & $M_{\odot}$ & $9.77 \pm 0.43$ \\
$\mu_{\rm gas}$ & $ $ & $0.49 \pm 0.12$ \\
$t_{\rm dep}$ & Gyr & $0.22 \pm 0.05$ \\
log$_{10}[n_{H}]$ (PDR) & cm$^{-3}$ & $4.80 \pm 0.39$ \\
log$_{10}[G_0]$ (PDR) & Habing & $2.83 \pm 0.26$ \\
$T_s$ & K & $186 \pm 116$ \\
$V_{\rm rot}$ & km/s & $148 \pm 33$ \\
$V_{\sigma}$ & km/s & $32 \pm 10$ \\
$V/\sigma$ &  & $4.6 \pm 1.7$ \\
$M_{\rm dyn,disk}$ & $M_{\odot}$ & $(1.39\pm0.31) \times 10^{10}$ \\
$Z$ & 12 + Log[O/H] & $7.96 \pm 0.44$ \\
$r_{1/2}$ (Sersic) & $^{\prime\prime}$ & $0.40$ \\
\enddata


\tablecomments{All quantities have been corrected for lensing magnification.}
\end{deluxetable}

\begin{figure*}[!htb]
\centering
\includegraphics[scale=0.75]{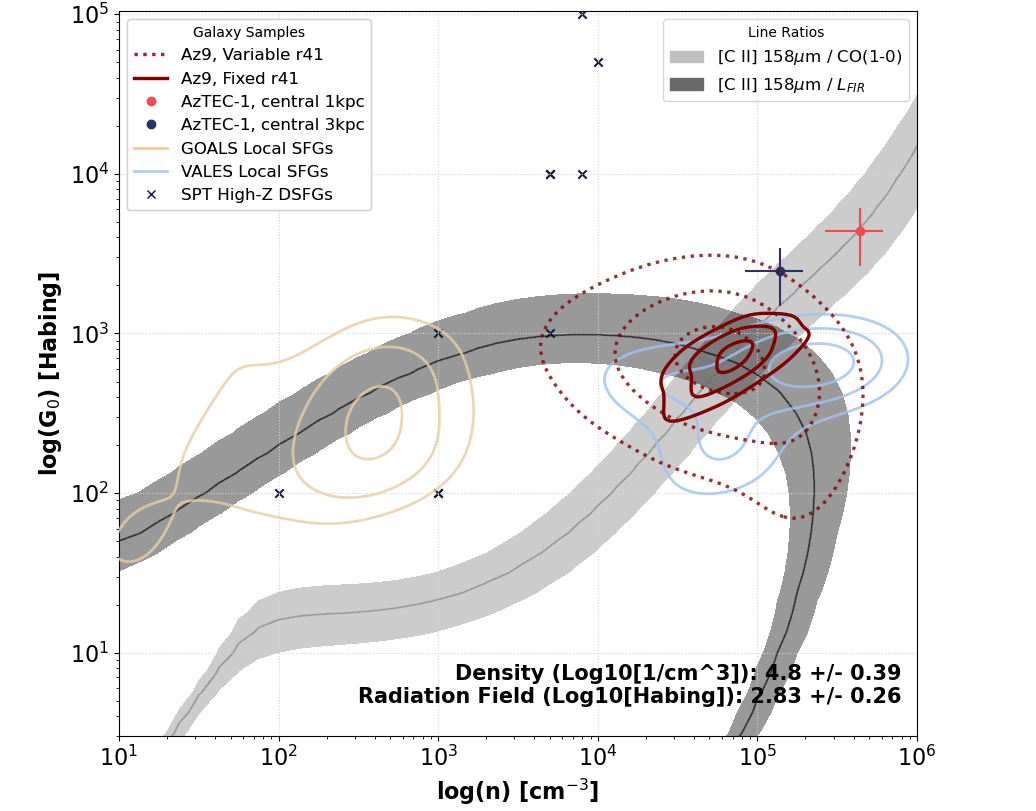}
\caption{Results of PDR modeling of Az9 (maroon contours) using the \citet{Wolfire1990,Wolfire2010} models, showing hydrogen density versus FUV radiation field strength. For Az9, the dotted contours represent our results when $r_{41}$ is drawn from a normal distribution with $r_{41} = 0.63 \pm 0.44$, while the solid contours represent the results when we used a fixed $r_{41} = 0.63$. The grey shaded areas represent the regions in parameter space allowed by the two observed line ratios. All contours are shown at 0.3, 0.6, and 0.9, levels. We overplot G and n for various other galaxy populations; GOALS LIRGs and sub-LIRGS \citep{DS2017}, VALES normal SFGs \citep{Hughes2017}, and high-z dusty star forming galaxies \citep{Gullberg2015}. We also show the location of AzTEC-1 \citep[both the inner 1 and 3 kpc][]{Tadaki2019}.
\label{fig:pdr}}
\end{figure*}

\subsection{Gas Conditions in PDRs}
In Figure 5, we show the results of our best-fit PDR model. We plot the position in radiation field / density space of Az9 alongside a sample of galaxies from the literature, including GOALS LIRGs and sub-LIRGS, \citep{DS2017}, normal SFGs \citep{Hughes2017}, and high-z dusty star forming galaxies \citep{Gullberg2015}. We also show AzTEC-1 \citep{Tadaki2019}, a galaxy at comparable redshift to Az9 but with a much higher mass and SFR. We note that, with the exception of the SPT sources, the literature points shown on this plot have been subject to identical corrections as those we apply to Az9 (i.e. doubling the CO luminosity and correcting for the [C II] PDR fraction). Additionally, the GOALS sources use [O I] 63 $\mu m$ as their dense gas tracer rather than CO. We show our results for Az9 (maroon) in the form of 2 sets of density contours: one set (dashed) representing the distribution of results produced assuming a variable $r_{41}$, and one set (solid) representing the results assuming a fixed $r_{41} =0.63$. Under the assumption of a fixed $r_{41} = 0.63$, we find that Az9 has a hydrogen density log[$n_H$ $cm^{-3}$] = $4.80 \pm 0.39$ and a radiation field strength log[G$_0$ Habing] = $2.83 \pm 0.26$. Using a normally distributed $r_{41} = 0.63 \pm 0.44$, we find log[$n_H$ $cm^{-3}$] = $4.70 \pm 0.72$ and log[G$_0$ Habing] = $2.73 \pm 0.91$

Regardless of how we handle $r_{41}$, Az9 appears to host PDR conditions consistent with the local normal star-forming galaxies studied in the VALES program \citet[][]{Hughes2017}, who derived PDR conditions in their sample using measurements similar to those used in our study. The PDR conditions in Az9 are less similar but still broadly comparable to those found in AzTEC-1, with a somewhat less intense radiation field  - sensible, given that AzTEC-1 is a massive ($M_{*} \sim 10^{11} M_{\odot}$) galaxy undergoing an aggressive starburst \cite[$SFR \sim 10^3 M_{\odot}/yr$,][]{Tadaki2018}. In fact, it is curious that these two systems, with star formation rate surface densities varying by a factor of $\sim 10$ ($\Sigma_{SFR}$ $\sim$ 2.5 $M_{\odot}$/yr/kpc$^2$ for Az9, \citet{Pope2023} and $\Sigma_{SFR}$ $\sim$ 270 $M_{\odot}$/yr/kpc$^2$ for AzTEC-1, \citet{Sharda2019}) have similar PDR conditions. Although it is in agreement in with the GOALS sources in radiation field terms, Az9 is well offset from them in density space, possibly due to the use of different PDR tracers.

The PDR models also estimate a surface temperature of $T_s = 186 \pm 116$ K in the variable $r_{41}$ case. This is an upper limit to the gas temperature in PDRs as the surface gas most directly exposed to radiation will be the hottest.

\subsection{Kinematic Properties}

\begin{figure*}[!htbp]
\vspace{0.1in}
\includegraphics[width=1.0\textwidth]{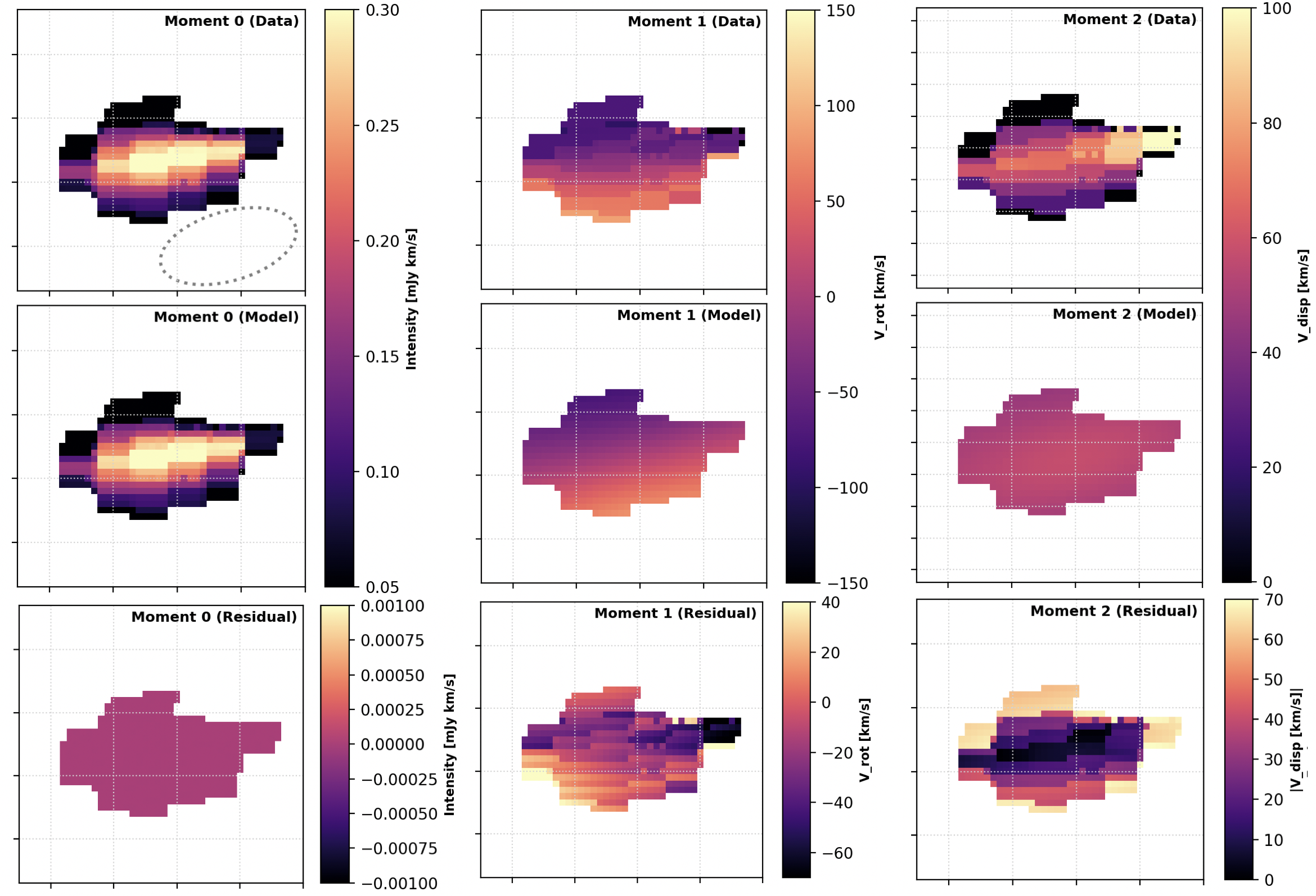}
\caption{Moment 0,1, and 2 maps of Az9 CO(4-3) emission are shown in columns 1, 2 and 3.
Kinematic data (top), model (middle), and residuals (bottom) produced by 3D-Barolo of our source-plane reconstruction of the CO emission are shown. The shape and orientation of the source-plane resolving beam is also shown in the first panel.}
\end{figure*}

We show the kinematic maps (moment 0, moment 1 and moment 2 in the data, model, and residuals) produced by 3D-Barolo in Figure 6. With about $\sim 2$ beams resolving Az9 along the axis of rotation (similar to the resolution achieved by \citet{Jones2021b,Pope2023}), Az9 is decently fit by a tilted-ring model and is consistent with a rotating system. 3D-Barolo also provides the best-fit velocity dispersion and circular velocity within each model ring. The distribution of these values as a function of ring outer radius (i.e. the rotation curve of our model of Az9) is presented in Figure 7. The shape of the rotation curve is broadly consistent with the shape of other rotation curves derived for rotating disks observed at low spatial resolution \cite[e.g.][]{Teodoro2015}. We find a mean circular velocity $V_{rot}$ = $148\pm32$ km/s in the outer 5 rings (where the velocity profile is approximately flat) and mean velocity dispersion $V_{\sigma}$ = $32\pm10$ km/s across the source. The value of $V/\sigma$ for CO-emitting gas in Az9 is $4.6 \pm 1.7$, implying that Az9 is a rotationally dominated system. This result is consistent with the $V/\sigma$ found for the [C II] $(5.3 \pm 3.6)$ in the same system \cite[][]{Pope2023}. These derived quantities are not sensitive within uncertainty to initial guesses of $V_{rot}$ and $V_{\sigma}$. In some cases, kinematic analyses performed at low spatial resolution can misclassify mergers as disks \citep{Rizzo2022}. However, following \citet{Pope2023}, the lack of a multiply-peaked velocity profile in our data lends credence to the disk interpretation.

\begin{figure}[!htb]
\vspace{0.1in}
\includegraphics[width=0.45\textwidth]{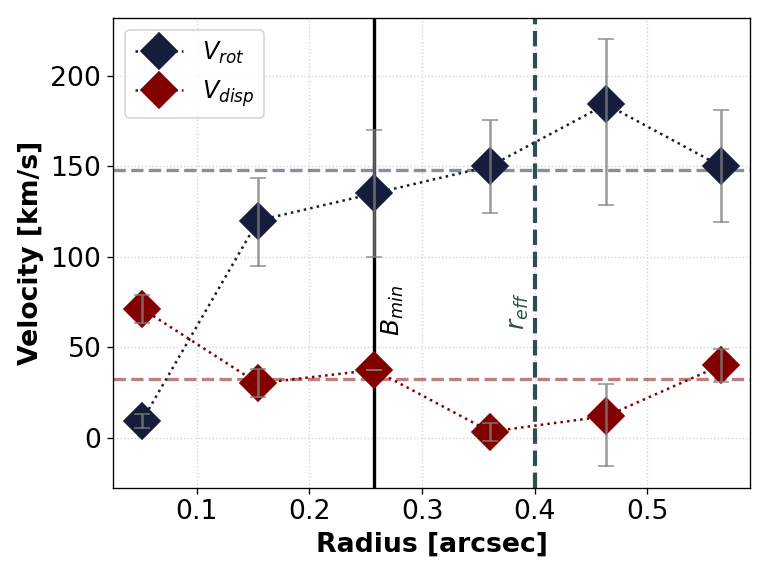}
\caption{Velocity radial profiles for CO.
The circular velocity and velocity dispersion of Az9 as a function of model ring central radius in differential bins. These represent the value in each ring in our 3D-Barolo tilted-ring model. A vertical solid line is placed at the scale of the source-plane beam minor axis, along which the axis of rotation is resolved. Horizontal dashed lines represent the radially averaged value of each component.}
\end{figure}

\subsubsection{Dynamical Mass}
In order to estimate the dynamical mass of Az9 assuming a disk-like geometry, we use an equation first developed by \citet{Wang2013} and commonly used throughout the literature \citep[e.g.][]{Capak2015, Cassata2020}: $M_{dyn, disk}[M_{\odot}] = 1.16 \times 10^{5} V_{cir}^2 D$ where $D$ is the disk diameter in kpc and $V_{cir}$ is the circular velocity of the gas disk in km$/$s. Using twice our half-light radius $R_{1/2}=2.74$ kpc as the diameter and our radially-averaged rotational velocity $V_{rot}$=148 $\pm$ 32 km/s, we find a dynamical mass of $M_{dyn,disk} = 1.39 \pm 0.31 \times$ 10$^{10}$ $M_{\odot}$.

Some systematic uncertainties must be considered when interpreting this value. Simulations of massive disk galaxies suggest that gradients in the turbulent pressure at large radii ($\gtrapprox 3$kpc) cause the observed circular velocity to fall off relative to the spherically symmetric zero-pressure expectation $V_{c} = \sqrt{G M_{enc} / r}$ \citep{Wellons2020}, which may bias dynamical masses low by up to $\sim40\%$, and is more important at higher masses and redshifts. Beam smearing, which causes the observed velocity gradient to be reduced relative to its intrinsic value, is common in low-resolution observations like those we present here. As it causes rotational velocity to be confused with velocity dispersion, it may reduce the derived $V/\sigma$ and dynamical mass if not handled properly \citep{Bukert2016,Wuyts2016,Molina2019}. We do not believe either phenomenon has a significant impact on our results; turbulent pressure support is only a major contributor to disk structure for thick disks or systems with a low $V/\sigma$ \citep{Molina2019}, and 3D-Barolo accounts for beam smearing through its convolution step \citep{Teodoro2015} and fully 3-D approach.

\subsubsection{Dynamical Constraints on $\alpha_{CO}$}
The highest $\alpha_{CO}$ possible for Az9 given its dynamical mass (assuming it contains \textit{no neutral gas or dark matter}, only stars and molecular gas, and using $r_{41} = 0.63$) is $\alpha_{CO} = 6.1 \pm 1.6$ $M_{\odot}$ (K km s$^{-1}$ pc$^2$)$^{-1}$. The true value of $\alpha_{CO}$ is likely lower than the value quoted above, as Az9 is certain to contain some amount of dark matter and neutral gas. As such, for the Monte Carlo gas mass analysis described in Section 3.8, we set the upper limit on $\alpha_{CO}$ to $6.1$. Although the neutral gas mass fraction is uncertain (and often neglected in this sort of analysis) a dark matter mass fraction on the order of $\sim 20 \%$ is commonly quoted in the literature \citep[e.g.][]{Hodge2012,Cassata2020}. Interestingly, our $\alpha_{CO}$ upper limit becomes consistent with the classical $\alpha_{CO} = 3.8$ used by \citet{Tacconi2018} for normal star-forming galaxies once adjusted down by $\sim 20 \%$. 

\subsection{Metallicity Constraints}

Alongside the existing measurement of the [C II] line in Az9, our new measurement of the [N II]205$\mu$m line allows us to estimate the gas-phase metallicity of this system. \cite{Croxall2017} find that the [C II]/[N II]205$\mu$m ratio can be used as a rough metallicity tracer and calibrate a fitting function on the \cite{Pilyugin2005} metallicity scale: \[  \textrm{12+log$_{10}$[O/H]}_{PT05} = 8.97 - 0.043 \times \textrm{[C II]} / \textrm{[N II}] \] Using this relation we find $\textrm{12+log$_{10}$[O/H]}_{PT05} = 7.96 \pm 0.44$. The high uncertainty here is driven by the low signal-to-noise of our observation of the [N II]205$\mu$m line. This is consistent with the metallicity one would estimate via the mass-metallicity relation at \textit{z}=4.3 \citep[$\textrm{12+log$_{10}$[O/H]}_{PT05} \sim 7.88$,][]{Genzel2015}.
Assuming this metallicity, we estimate the $\alpha_{CO}$ using the recipe developed by \citet{Genzel2015}: \[ \alpha_{CO}^Z = \alpha_{CO,MW} \times 10^{-1.27(12+\log(O/H)_{PP04}-8.67)} \] where $\alpha_{CO,MW}$ is the Milky Way CO-to-H$_{2}$ conversion factor including a correction factor for helium, typically taken to be 4.36 $M_{\odot}$ $/$ K km s$^{-1}$ pc$^2$ \citep{Strong1996}. We obtain the very high value of $\alpha_{CO} = 34$ $M_{\odot}$ / K km s$^{-1}$ pc$^2$ for Az9 using this method. Using this $\alpha_{CO}$ and the \cite{Castillo2023} value $r_{41}=0.63$ for the 4-to-1 conversion factor, we calculate a molecular gas mass of $M_{H2}$ = $8.2 \pm 1.2$ $ \times 10^{10}$ $M_{\odot}$. This is unphysical, as it is much higher than the total dynamical mass of the system regardless of the assumed geometry. Additionally, assuming $\alpha_{CO} = 34$, an extremely high $r_{41} = 3.7$ would be required in order to produce a molecular hydrogen mass equal to the dynamical mass. The result that locally calibrated fitting functions for $\alpha_{CO}$ may not be appropriate in high-redshift galaxies is not new \cite[e.g.][]{Narayanan2012,DZ2017} and this Az9 result confirms that more work must be done to understand the conversion factor in high z, low mass galaxies.

\section{Discussion} \label{sec:results}

Combining our new CO (tracing dense molecular gas) and [N II]205$\mu$m (an important cooling line for ionized regions) measurements of Az9 with existing measurements of the [C II] (an important cooling line for PDRs) provides us with insight into the nature of multiple gas phases within a low mass dusty galaxy at $z=4.27$. Our observations paint a picture of a system that - despite its high obscured SFR - is surprisingly ``normal", with PDR conditions comparable to local normal dusty galaxies and kinematics suggesting a dynamically cold disk. A system with ``normal"  kinematics and ISM properties like Az9 should be able to sustain them for an extended period, which perhaps suggests that Az9 is representative of a larger low-mass galaxy population heretofore inaccessible without the benefit of gravitational lensing.

\begin{figure}[!htb]
\includegraphics[width=0.45\textwidth]{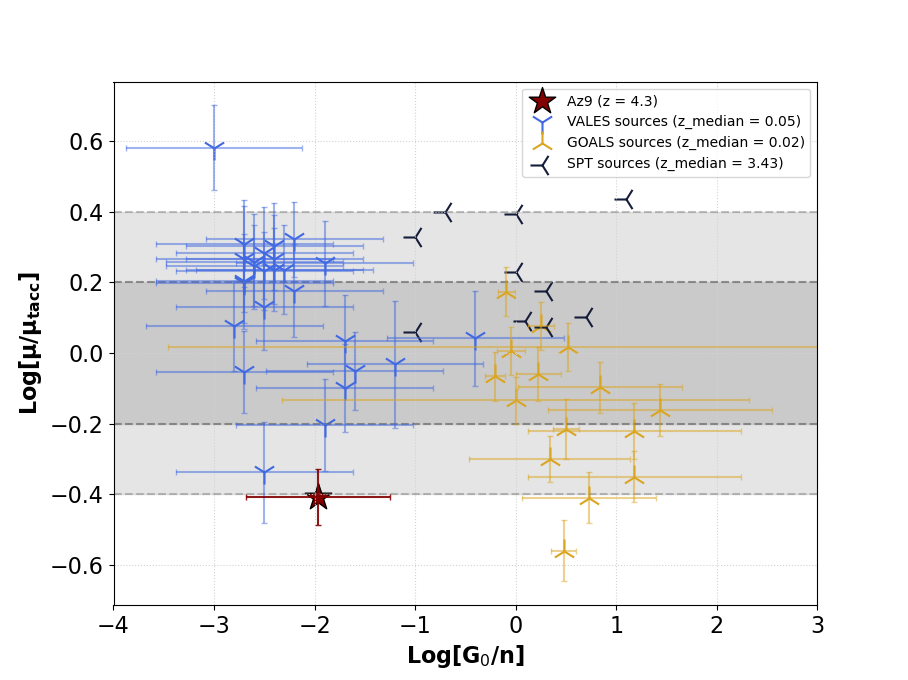}
\caption{We show the logarithmic distance from the \cite{Tacconi2010} molecular gas mass to stellar mass fraction sequence as a function of ISM conditions, log[$G_0$/$n$]. The shaded regions corresponds to $1 \sigma$ and $2 \sigma$ distances from the \citet[][]{Tacconi2010} relation. We also show 3 samples from the literature: VALES \cite[][]{Villanueva2017,Hughes2017}, GOALS  \cite[data from][]{Armus2009,DiazSantos2017,U2012,Yamashita2017}, and a subset of the SPT sources \citep{Gullberg2015}. An $\alpha_{CO}$ = 3.6 $M_{\odot}$ (K km s$^{-1}$ pc$^2$)$^{-1}$ has been adopted for all literature points. In the nearby galaxy samples, there is a loose trend wherein systems with a higher log$_{10}$[$G_0$/$n$] correspond to lower molecular gas mass to stellar mass fraction. Az9 is unremarkable in its position on the log$_{10}$[$G_0$/$n$] axis and appears somewhat gas-poor, though the significant systematic uncertainties in molecular gas content are suppressed in this plot.}
\label{fig:delta_fgas}
\end{figure}

\subsection{PDR Conditions: Evidence of a ``Normal" System?}

We can gain some additional insight into how Az9 compares to other populations by looking at the molecular gas content of Az9 relative to the \citet{Tacconi2018} relation, log$_{10}$[$\mu_{H_2}$/$\mu_{H_2,tacc}$], in the context of other ISM properties. In Figure 7, we show log$_{10}$[$\mu_{H_2}$/$\mu_{H_2,tacc}$] as a function of log$_{10}$[$G_0$/$n_H$] for Az9 alongside the same sample of galaxies we showed in Figure 4. Az9 has ISM conditions and molecular gas mass to stellar mass fraction in log$_{10}$[$G_0$/$n_H$] space that are consistent with VALES \citep{Jones2021b}. This suggests that a relatively ``normal'' $\alpha_{CO}$ is appropriate: if for example we had found that Az9 possessed an extremely strong internal radiation field, radiation would be able to drive photodissociation deeper into a given component cloud and it would make sense to use a higher $\alpha_{CO}$. However, in terms of its PDR conditions, Az9 appears broadly similar to local universe dusty star-forming galaxies - a comparable $\alpha_{CO}$ is probably reasonable.

The PDR conditions of Az9 are not only interesting in the context of $\alpha_{CO}$ - they also represent the first time these properties have been measured for a galaxy with log[$M_{*} (M_{\odot})$] $\lesssim 10$ at high redshift. Az9 appears to possess PDR physical conditions quite similar to those observed in a sample of local-universe dusty star forming galaxies, despite those galaxies hosting somewhat higher stellar masses (log$_{10}[M_{*} (M_{\odot})] = 9.79 - 11.2$ compared to log$_{10}[M_{*} (M_{\odot})] = 9.33$ in Az9) and significantly lower specific star formation rates (sSFR(Gyr$^{-1}$) $= 0.05 - 4.03$ compared to sSFR(Gyr$^{-1}$) $= 14.15$ in Az9) \citep{Hughes2017, Villanueva2017}.  

One explanation for the surprising similarity between Az9 and the VALES local-universe sources is a picture in which the nature of the ISM in individual dusty star-forming clumps is broadly similar in all systems undergoing ``normal" star formation. \citet{DiazSantos2017} suggest that the youngest PDRs account for the highest G/n ratios. Young PDRs host younger stars which produce more intense radiation radiation fields, and are likely more compact as they have had less time to expel their dusty envelopes. In this model, the high G/n ratios typical of starbursts are due to the presence of many such PDRs, individually short-lived but constantly replenished as they are triggered by mergers. 

In contrast, the more typical PDR conditions of Az9 suggest a more gentle, extended history of dust-obscured star formation rather than a recent or ongoing starburst. This in turn suggests that the normal dust-obscured star forming phase is reasonably long for galaxies like Az9. If so, Az9 could be representative of a more widespread population of low-mass main-sequence systems - with an intrinsic 1.1mm flux of merely $S_{1.1mm} \sim 0.1$ mJy \citep{Pope2017}, any similar unlensed systems would be well below the detection threshold of all existing blank-field single-dish surveys \citep[e.g.][]{Hodge2013,Geach2017,Stach2019} and all but the deepest (and therefore smallest) interferometric \citep[e.g.][]{Franco2018} surveys. We can further test the idea that systems like Az9 are widespread with deep observations from LMT/TolTEC.

\subsection{Dust Content}

We derive a dust mass upper limit of log$_{10}[M_{\rm dust}] < 7.73$, consistent with the dust mass estimated via SED fitting in \citet{Pope2023} as well as expectations from semi-analytical models \citep{Popping2017}. The gas-to-dust ratio (GDR) is consistent with GDRs seen in local universe galaxies of comparable metallicity; we find Az9 has a GDR lower limit of log$_{10}[GDR_{Az9,lower}] \geq 2.04$ compared to the expected (assuming the low-metallicity branch) log$_{10}[GDR_{expected}] = 2.92 \pm 1.28$ \citep{Remy2014}.

It is also worthwhile to consider how Az9 is so heavily obscured so early on in the history of the universe. At high redshift, condensation of metals on to existing small grains in the cold neutral medium (CNM) or molecular clouds is likely to be the predominant dust formation channel \citep{Draine2009,Michalowski2015,Popping2017}. The interstellar dust population is also dominated by grains formed in molecular clouds in systems like the Milky Way \citep{Zhukovska2008}. Using a common expression for the dust accretion timescale in the CNM \citep[e.g.][]{Asano2013,Triani2020}: 

\begin{equation}
\label{eq:tau_acc}
\tau_{\rm acc} =  2 \times 10^7 \times \bigg(\frac{n_{\rm mol}}{100
  \,\rm{cm}^{-3}}\bigg)^{-1}\,\bigg(\frac{T_{dust}}{50\,\rm{K}}\bigg)^{-1/2}\,\bigg(\frac{Z}{Z_{\odot}}\bigg)^{-1} \text{[yr]}
\end{equation}

and the $H_{2}$ density we found in our PDR model $n_{mol} \sim 10^{4.8}$ cm$^{-3}$, $T_{dust} = 40$K, and our metallicity 12 + log$_{10}$[O/H] $\sim$ $8$, we find Az9 has a dust growth timescale $\tau_{acc} \approx 0.2$ Myr. This is quite rapid, but timescales on the order of $10^5$yr is reasonable some situations, such as the depletion of metal ions onto small grains in normal HI clouds \citep[][]{Draine2004,Draine2009}. The majority of this growth is expected to occur on the surface of small grains, as these grains provide a significant fraction of the dust total surface area and are likely to be negatively charged \citep[][]{Draine2009}. Coagulation of small grains may also be an important process in dense regions of the ISM \citep{Hirashita2023}. 
Although both our PDR results and this dust coagulation timescale are uncertain, we can conclude qualitatively that the presence of dense PDRs in Az9 may allow for the rapid assembly of dust mass in these small dense regions.

With a dynamical timescale of roughly $\sim30$Myr (calculated as $R/V$), this suggests that in its current state, Az9 is able to assemble dust mass in its densest ISM regions more rapidly than that dust can be transported throughout its structure. Thus, we suggest that the high obscuration fraction of Az9 might be due to the presence of clumpy dusty regions strongly associated with dense gas and ongoing star formation. This is an idea that could be tested with future high-resolution ALMA data. 

\section{Summary} \label{sec:highlight}

In this work, we present new ALMA observation of the CO(4-3) and [N II] lines in the z=4.273 low-mass star-forming galaxy MACSJ0717\_Az9. The main conclusions of this paper are as follows: 
\begin{itemize}
\item We find that Az9 is moderately deficient in molecular gas compared to the existing molecular gas mass to stellar mass fraction scaling relation, with a molecular gas mass of log$_{10}[M_{H_{2}} ] = 9.77 \pm 0.43$ and molecular gas mass to stellar mass fraction $\mu_{\rm gas} = 0.49 \pm 0.12$. This is consistent with other low-mass systems observed at high redshift. The strength of this conclusion is limited by the systematic uncertainties on $\alpha_{CO}$ and $r_{41}$, which are poorly unconstrained at high redshift and low stellar mass.
\item Based on observations of the CO(4-3) line, we find that Az9 is a rotationally-dominated system, with $V_{rot} = 148 \pm 33$ km/s and $V_{disp} = 32 \pm 10$ km/s. These observations trace a different gas phase than previous work and provide strong evidence that Az9 hosts a stable molecular disk.
\item Az9 has PDRdd conditions similar to local normal dusty star-forming galaxies, with a mean hydrogen density log$_{10}$[$n_H$ $cm^{-3}$] = $4.80 \pm 0.39$ and a mean radiation field strength log$_{10}$[G$_0$ Habing] = $2.83 \pm 0.26$ in these regions. This is the first measurement of PDR conditions in a galaxy with log$_{10}$[$M_{*} (M_{\odot})$] $< 10 $ at z $>$ 4.
\item Assuming an $r_{41}$ of 0.63, we find that the dynamical mass favors an upper limit on $\alpha_{CO}$ of 6.1 assuming a disk-like geometry. This suggests that $\alpha_{CO}$ values based on extrapolating scaling relations from more massive, local galaxies may not be appropriate for galaxies like Az9.
\item Based on our new observations of the [N II] 205 $\mu$m line, we estimate the metallicity of Az9 to be 12 + log$_{10}$[O/H]$_{PTO5}$ = 7.96 $\pm$ 0.44. The high uncertainty on this value is driven by the low signal-to-noise of our [N II] 205 $\mu$m data.
\item We find that Az9 has a short dust growth timescale. This implies that Az9 is able to assemble dust mass in its ISM more rapidly than that dust can be transported throughout its structure, resulting in a clumpy dust geometry. It implies that any metals injected into its dense ISM are very quickly depleted onto the surface of dust grains.
\end{itemize}

In most respects, Az9 appears to be a fairly normal low-mass star forming galaxy for its redshift. Az9 is mainly notable for two reasons: (1) it has a high obscured star formation fraction, and (2) it appears to be have molecular gas content somewhat below what might be expected based on extrapolating existing scaling relations. Deeper observations of dust continuum emission in the Rayleigh-Jeans tail and a stronger constraint on the metallicity of Az9 through rest-frame optical lines would be excellent next steps, allowing us to understand how Az9 assembled its metal content in significantly more detail.

Az9 hosts a dynamically cold, stable disk and has ISM conditions comparable to local normal star-forming galaxies, suggesting that it can maintain its current dust-obscured phase for an extended period. If so, this suggests that Az9 may be representative of a more numerous population of low-mass dust-obscured galaxies at $z\sim4$. Future mm/sub-mm surveys with $3\sigma$ depths better than $\sim$0.1 mJy such as the LMT/TolTEC UDS survey \citep{Montana2019} may be able to detect more systems like Az9, though with current instruments gravitational lensing is the only practical way to study these systems in detail. 

\begin{acknowledgments}
The authors would like to thank the NAASC data reduction training team, who provided valuable assistance during the data reduction process. This paper makes use of the following ALMA data: ADS/JAO.ALMA\#2021.1.00272.S. ALMA is a partnership of ESO (representing its member states), NSF (USA) and NINS (Japan), together with NRC (Canada), MOST and ASIAA (Taiwan), and KASI (Republic of Korea), in cooperation with the Republic of Chile. The Joint ALMA Observatory is operated by ESO, AUI/NRAO and NAOJ. The National Radio Astronomy Observatory is a facility of the National Science Foundation operated under cooperative agreement by Associated Universities, Inc.
\end{acknowledgments}

%

\vspace{5mm}
\facilities{ALMA, HST(WFC3), LMT(AzTEC)}


\software{astropy \citep{astropy1},  
          }



\bibliography{sample631}{}
\bibliographystyle{aasjournal}



\end{document}